%% file: main.tex
\documentclass[11pt]{article}
\usepackage[preprint]{acl}
\usepackage{times}
\usepackage{latexsym}
\usepackage[T1]{fontenc}
\usepackage[utf8]{inputenc}
\usepackage{microtype}
\usepackage{inconsolata}
\usepackage{graphicx}
\usepackage{amsmath}
\usepackage{amssymb}
\usepackage{booktabs}
\usepackage{array}
\usepackage{xcolor}
\DeclareMathOperator{\Top}{Top}
\DeclareMathOperator{\sg}{sg}

\title{Pix2Evidence: A Multimodal RAG System for Text and Visual Evidence}
\author{Ziniu Liu$^{1,2}$ \quad  Xiaoyang Chen$^{2}$ \quad  Dezhi Ye$^{2*}$ \quad  Jie Liu$^{2}$ \quad   Jin Ma$^{2}$ \quad  Hao Deng$^{1}$\thanks{Corresponding author}\\
$^{1}$Tongji University \quad \quad  $^{2}$Tencent Yuanbao\\
{\tt\small denghao1984@tongji.edu.cn \quad dezhiye@tencent.com}}
\begin{document}
\maketitle

\begin{abstract}
Retrieval-augmented generation (RAG) over visually rich documents must
find both relevant content and the evidence that supports an answer.
Screenshot retrieval preserves images and page structure, but written
facts within screenshots also need to be searchable.
Moreover, topical relevance alone does not identify which retrieved
tile contains evidence for the answer.
We introduce Pix2Evidence, a multimodal RAG system that encodes each
screenshot tile together with its available transcription into a
single fused image--text representation, so that layout and readable
text are considered jointly during retrieval.
To guide evidence selection, a vision-language teacher describes the
visible content that supports each training question. 
We use these descriptions to adapt the retriever, then rerank the
retrieved candidates before answer generation. Extensive experiments demonstrate that Pix2Evidence outperforms state-of-the-art methods. 
Across five widely-used Wikipedia QA benchmarks, the non-reranked Pix2Evidence leads the main
comparison in answer accuracy on four.
On MMSearch, it reaches 26.0\%, 4.4 percentage points above PixelRAG.
We will release the code and share derived data where licensing
and privacy reviews permit.
\end{abstract}

\section{Introduction}
Retrieval-augmented generation (RAG) grounds answers in an external
document collection \citep{realm,rag}.
A retriever selects candidate sources, and a reader generates an answer
from the material it receives \citep{dpr,fid}.
Open-domain QA often uses web content: Natural Questions pairs queries
with Wikipedia pages \citep{nq}, while WebQA draws on both visual and
textual web sources \citep{webqa}.

Vision-based retrieval can preserve information that is difficult to
capture as plain text, including images, tables, and page layout.
Document image retrievers represent rendered pages rather than relying
only on extracted text \citep{dse,colpali,hybrid_visual}.
PixelRAG applies this approach to Wikipedia QA by indexing screenshot
tiles at scale \citep{pixelrag}.
However, many questions still depend on written facts such as names,
dates, and numbers.
When these facts are embedded in an image, a retriever must recognise
the text visually before it can match it to the question.
This motivates encoding the text together with the visual tile in a
single fused representation, so that written facts become directly
available for retrieval without discarding visual and layout cues.

Retrieving relevant content is only the first step.
A tile may match a question's topic without containing the fact needed
for the answer, especially when the reader sees only a few tiles \citep{realign}.
Standard question and document supervision \citep{ance} identifies a relevant tile
but does not directly describe the content within it that supports the
answer.
We therefore generate question-conditioned descriptions of visible
answer evidence during training.
A vision-language teacher uses the question, a positive tile, and the
reference answer to describe the supporting content.
These descriptions provide evidence-focused supervision for adapting
the retriever.

\begin{figure*}[t]
\centering
\includegraphics[width=\textwidth]{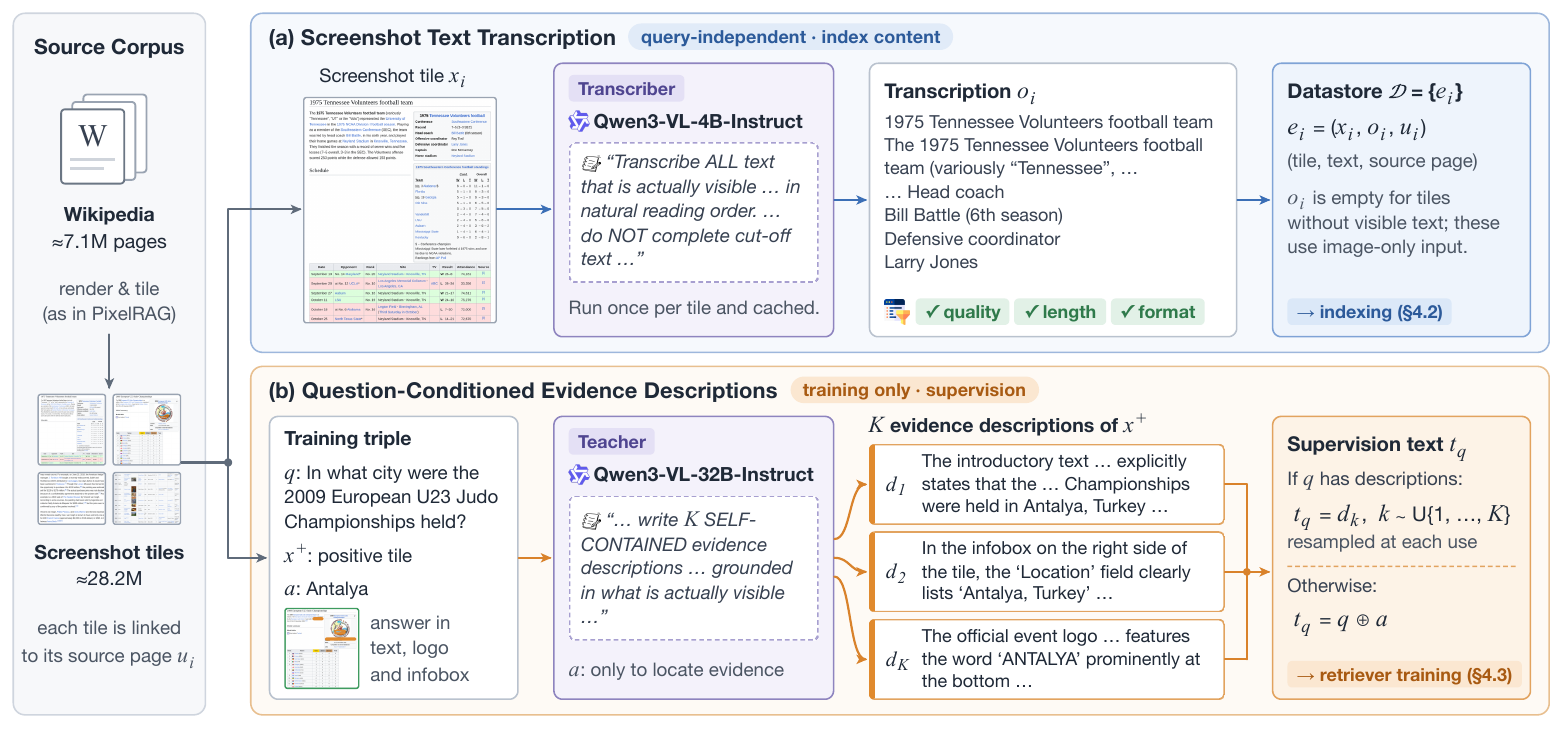}
\caption{An overview of our data construction pipeline, which consists
of two parts. (1) Screenshot Text Transcription: Text is extracted and
transcribed from image patches using Qwen3-VL-4B-Instruct.
(2) Question-Conditioned Evidence Descriptions: A high-capacity model
serves as a teacher to generate evidence description text, facilitating
alignment during training.}
\label{fig:evidence-generation}
\end{figure*}

\begin{figure*}[t]
\centering
\includegraphics[width=\textwidth]{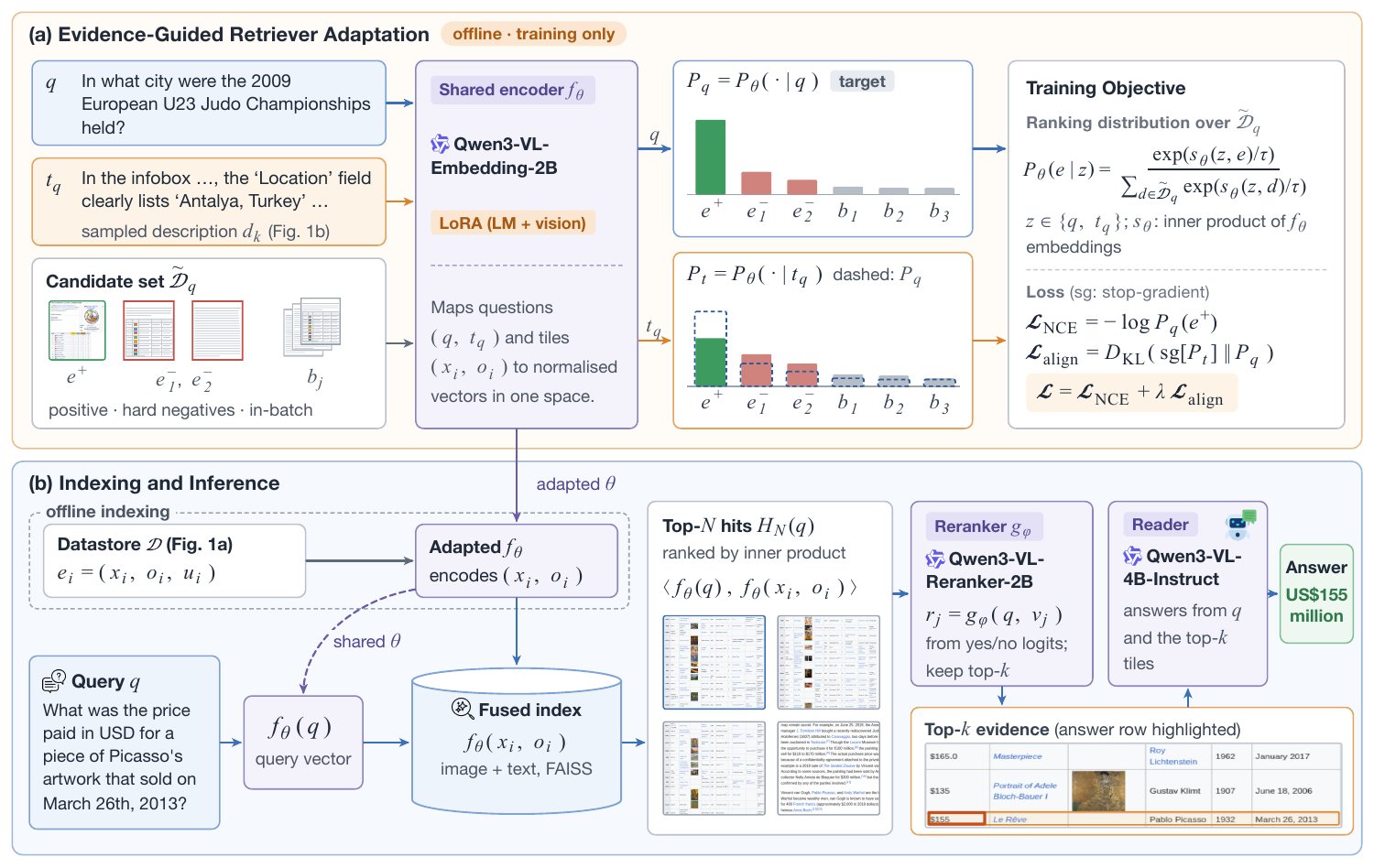}
\caption{Overview of Pix2Evidence. Top: training candidate set for
question $q$ with positive, negative, and in-batch tiles; dashed
distributions align question ranking $P_q$ to evidence ranking $P_t$.
Bottom: tiles $e_i=(x_i,o_i,u_i)$ are encoded by $f_\theta$ into a
fused image--text index; query $q$ retrieves top-$N$ candidates, a
vision-language reranker selects top-$k$, and a reader answers. The
example's highlighted row gives US\$155 million.}
\label{fig:overview}
\end{figure*}

Pix2Evidence addresses these two needs with fused multimodal retrieval
and evidence-guided training.
Our retrieval component integrates textual and visual information: it
jointly encodes each screenshot and its available transcription into a
shared multimodal representation, retrieves candidates from this fused
view, and reranks them before passing them to the reader.
During training, the teacher descriptions make the supporting evidence
explicit.
We use them to align the candidate rankings induced by a question and
its evidence description, alongside contrastive learning.
In this way, retriever adaptation attends not only to which tile is
relevant, but also to the evidence that makes it useful for answering.
Figure~\ref{fig:overview} illustrates the inference setting;
Sections~\ref{sec:data} and~\ref{sec:framework} detail the data and
training procedure.

We evaluate the complete system on five Wikipedia QA benchmarks over
approximately 7.1 million pages and 28.2 million screenshot tiles.
Our non-reranked configuration numerically exceeds PixelRAG in answer
accuracy on four benchmarks.
On MMSearch it reaches 26.0\%, an improvement of 4.4 percentage points
over PixelRAG.
Additional configurations examine how the input representation,
candidate fusion, and reader depth affect these outcomes.
Our contributions are:
\begin{itemize}
  \item Starting from PixelRAG's rendered screenshot tiles, we create
  transcriptions for indexing and teacher descriptions of
  answer evidence for retriever training.
    \item We develop a multimodal RAG system that fuses image and text
  into a single retrievable representation and adapts its encoder with
  evidence-guided training.
  \item We evaluate the system across five QA benchmarks and multiple
  configurations. Its non-reranked configuration scores above PixelRAG on four and exceeds both text-extraction baselines on all five.
\end{itemize}

\section{Related Work}
\paragraph{Retrieval-augmented generation.}
REALM and RAG incorporate external retrieval into language modelling
and knowledge-intensive prediction \citep{realm,rag}.
DPR learns dense question--passage representations \citep{dpr}, and
Fusion-in-Decoder combines multiple passages during generation
\citep{fid}.
Beyond initial retrieval, instruction-following language models can
rerank candidate passages, as demonstrated by RankGPT \citep{rankgpt}.
Readers do not always handle distracting retrieved contexts reliably
\citep{implicit_robustness}.
We study this selection problem in multimodal RAG, where text and
visual content may offer different routes to the same answer.

\paragraph{Visual documents and screenshot retrieval.}
Document understanding models combine text, layout, and image features,
as in LayoutLMv3 \citep{layoutlmv3}, or generate directly from document
images without a separate OCR module, as in Donut \citep{donut}.
WebSRC pairs page screenshots with HTML and questions that require
structural understanding \citep{websrc}.
VisualMRC studies answer generation from document images collected
across web domains \citep{visualmrc}.
These tasks show why reading web content involves both text and visual
organisation, although neither evaluates our web-scale retrieval setup.
For retrieval, DSE encodes screenshots into dense representations
\citep{dse}; ColPali extends late interaction from ColBERT to visual
pages \citep{colbert,colpali}.
VisRAG retrieves document images for vision-language generation
\citep{hybrid_visual}, and PixelRAG scales screenshot retrieval to web
QA \citep{pixelrag}.
SERVAL indexes model-generated page descriptions \citep{serval}.
We instead index transcribed text from screenshot tiles and reserve
question-conditioned descriptions for training.
ViDoRe V3 evaluates retrieval, grounding, and generation in visually
rich documents \citep{vidorev3}.
Our setting keeps web tiles as retrieval units and studies how both
textual and visual views affect evidence selection.

\paragraph{Hybrid retrieval and multimodal evidence fusion.}
WebQA asks systems to select and combine visual and textual web sources
\citep{webqa}.
MultiModalQA studies questions that require reasoning across text,
tables, and images \citep{multimodalqa}.
Both illustrate the need for evidence beyond a single text passage,
but they use different task settings from our Wikipedia tile index.
MuRAG retrieves from a multimodal memory \citep{murag}, while VisRAG
passes retrieved document images to a vision language generator
\citep{hybrid_visual}.
In our system, the retriever encodes each tile together with its
available transcription in one fused representation, so that readable
text and layout are considered jointly rather than in separate indexes.

\paragraph{Evidence-guided retrieval supervision.}
Contrastive learning distinguishes positive documents from negatives
\citep{infonce,dpr}.
RocketQA improves dense retrieval with additional negatives and data
augmentation \citep{rocketqa}.
Other work uses generated supervision: GPL combines synthetic queries
with relevance labels from a cross encoder to adapt dense retrieval
\citep{gpl}.
ReAlign generates question-focused descriptions for selected regions
of a document page, then aligns document rankings induced by the
question and those descriptions \citep{realign}.
We keep a distributional alignment objective but change what the
supervision is attached to.
Our teacher describes evidence inside a fixed screenshot tile, using the
training question and reference answer as guidance, and these
descriptions supervise the same encoder that produces our fused tile
representation.
Unlike ReAlign's region selection, our pipeline does not predict
bounding boxes.

\section{Data Construction}
\label{sec:data}
We use the Wikipedia screenshots already rendered and tiled by
PixelRAG \citep{pixelrag} as our visual corpus.
The corpus used here comprises approximately 7.1 million pages and
28.2 million tiles, each linked to its source page.
On top of these images, we construct two distinct kinds of text.
As shown in Fig~\ref{fig:evidence-generation}, query-independent transcriptions serve as index content, whereas
query-focused evidence descriptions provide training supervision.
The latter depend on a training question and reference answer and are
not available as test-time document fields.
Appendix~\ref{sec:corpus-details} reports corpus counts.

\subsection{Screenshot Text Transcription}
\label{sec:ocr}
We process every indexed tile $x_i$ with Qwen3-VL-4B-Instruct
\citep{qwen3vl} to transcribe visible text as $o_i$ in reading order.
The prompt discourages completion of words cut off at tile boundaries.
Transcriptions are cached independently of retrieval questions and
reference answers. We then perform quality, length, and format checks on the resulting transcriptions. 
About 85\% of indexed tiles have nonempty transcriptions; tiles with
no visible text retain an empty text field and use image-only inputs.

\subsection{Question-Conditioned Evidence Descriptions}
\label{sec:evidence}
Inspired by ReAlign \citep{realign}, we use
Qwen3-VL-32B-Instruct \citep{qwen3vl} as a teacher to generate $K$
descriptions of the evidence that supports the question for each training triple $(q,x^+,a)$.
The question defines what information is needed; the reference answer
$a$ helps the teacher locate that information in the positive tile $x^+$.
The generation instruction asks for supporting content visible in
the tile, rather than an unsupported restatement of the answer.
The resulting descriptions supply answer-conditioned training signals
intended to focus the multimodal retriever on useful evidence.
It differs from ReAlign's region descriptions \citep{realign} because
we describe fixed screenshot tiles and do not select bounding boxes.

Every training question has a supervision text $t_q$.
When generated descriptions are available, we sample one uniformly at
each training use; otherwise, we concatenate the question and its
reference answer to form $t_q$.
The reference answer helps the teacher locate supporting content
during data construction.
This supervision is used only for training and is not introduced at
inference.

\section{Pix2Evidence Framework}
\label{sec:framework}

\subsection{Overview}
\label{sec:overview}
Let $\mathcal D=\{e_i\}$ with $e_i=(x_i,o_i,u_i)$ denote a screenshot
tile, its transcription, and its source identifier.
Each tile is encoded together with its transcription into a fused
image--text representation. An overview of Pix2Evidence is shown in Fig~\ref{fig:overview}.

At inference, Pix2Evidence retrieves the top-$N$ candidates from this
fused index (Section~\ref{sec:retrieval}), reranks them with a
vision-language reranker (Section~\ref{sec:rerank}), and passes the
top-$k$ tiles to the reader.
The retriever is trained offline with evidence-guided supervision: a
contrastive term pulls the question toward its positive tile
(Section~\ref{sec:alignment}), and an alignment term makes the
question-side ranking follow the ranking induced by an evidence
supervision text (Section~\ref{sec:objective}).
Evidence descriptions are used only during training; inference relies
solely on the question and indexed document content.

\begin{table*}[t]
\centering
\small
\setlength{\tabcolsep}{2.5pt}
\begin{tabular*}{\textwidth}{@{\extracolsep{\fill}}lrrrrrrrrrr@{}}
\toprule
& \multicolumn{2}{c}{NQ} & \multicolumn{2}{c}{NQ-Tables} & \multicolumn{2}{c}{SimpleQA} & \multicolumn{2}{c}{MMSearch} & \multicolumn{2}{c}{EVQA} \\
\cmidrule(lr){2-3} \cmidrule(lr){4-5} \cmidrule(lr){6-7} \cmidrule(lr){8-9} \cmidrule(lr){10-11}
Method & Recall & Acc & Recall & Acc & Recall & Acc & Recall & Acc & Recall & Acc \\
\midrule   
No retrieval &  & 13.1 & & 10.9 & & 6.4 & & 13.5 & & 27.3 \\
\midrule
\multicolumn{11}{@{}l}{\textit{Text-based retrieval}} \\  
\quad mwparserfromhell & 48.6 & 27.7 & 35.0 & 18.3 & 72.9 & 58.4 & -- & 12.9 & 3.2 & 30.7 \\
\quad Trafilatura & 45.8 & 29.4 & 37.7 &  21.9 & 79.3 &  69.0 & -- & 13.5 & 3.2 & 29.2 \\
\midrule
\multicolumn{11}{@{}l}{\textit{Pixel-based retrieval}} \\
\quad PixelRAG (base) & 53.5 & 31.0 & 45.5 & 24.3 & 79.9 &  70.3 & -- & 19.9 & 17.0 & 39.4 \\
\quad PixelRAG & 58.9 & 31.1 & 50.7 & 25.8 & 83.1 & 72.8 & -- & 21.6 & \textbf{33.2} & 40.1 \\
\midrule
\multicolumn{11}{@{}l}{\textit{Fused retrieval}} \\ 
\quad Pix2Evidence  & \textbf{60.6} & 32.1 & \textbf{56.8} & 25.7 & 86.0 &  73.4 & -- & \textbf{26.0} & 30.8 & \textbf{40.6} \\  
\quad  Pix2Evidence (+reranker) & 57.8 & \textbf{33.2} & 55.8 & \textbf{27.1} & \textbf{86.9} & \textbf{77.6} & -- & 23.0 & 33.0 & 40.3 \\
\bottomrule
\end{tabular*}
\caption{Main results on five Wikipedia QA benchmarks. We use LLM-judged for SimpleQA, MMSearch, EVQA; EM for NQ and NQ-Tables.
MMSearch has no gold articles, so its recall is not reported. Bold marks
the best value in each column.}
\label{tab:main}
\end{table*}

\subsection{Inference Pipeline}
\label{sec:inference}

\subsubsection{Multimodal Retrieval}
\label{sec:retrieval}
We adapt Qwen3-VL-Embedding-2B \citep{qwen3vlretrieval} with LoRA
\citep{lora} in its language and vision components.
For tile $e_i=(x_i,o_i,u_i)$, $x_i$ is its screenshot, $o_i$ is its
transcription, and $u_i$ identifies the source page.
A shared encoder $f_\theta$ maps an input to a normalised vector;
$\theta$ denotes its adapted parameters.
We apply query and document input templates according to their roles.
The notation $f_\theta(q)$ is the vector for question $q$, and
$f_\theta(x_i,o_i)$ encodes the image with its available text.
If $o_i$ is empty, the input contains only $x_i$.

The retrieval score is
\begin{equation}
s(q,e_i)=\langle f_\theta(q),f_\theta(x_i,o_i)\rangle,
\label{eq:score}
\end{equation}
where angle brackets denote a vector inner product.
A FAISS index supports search over these normalised vectors
\citep{faiss}.
Appendix~\ref{sec:input-prompts} describes the input roles and the
availability of their prompt templates.

For a question $q$, let $H_N(q)$ be the ranked list of at most $N$
hits returned by the fused retriever:
\begin{equation}
\begin{aligned}
H_N(q)&=\Top_N^{\mathrm{list}}(s(q,\cdot)),\\
\mathcal C(q)&=H_N(q),\\
|\mathcal C(q)|&\leq N.
\end{aligned}
\label{eq:candidates}
\end{equation}
The operator $\Top_N^{\mathrm{list}}$ returns the highest-scoring
hits in rank order.
$|\mathcal C(q)|$ counts hit positions in the resulting list.
We pass $\mathcal C(q)$ to the reranking stage;
Appendix~\ref{sec:candidate-details} specifies candidate input
formats.

\subsubsection{Cross-Modal Reranking}
\label{sec:rerank}
Qwen3-VL-Reranker-2B \citep{qwen3vlretrieval} jointly scores each
question--candidate pair without task-specific reranker training.
Given the candidate list $\mathcal C(q)$ from the fused retrieval
stage, we compute for each candidate input $v_j$
\begin{equation}
\begin{aligned}
r_j&=g_\phi(q,v_j),\\
I_k(q)&=\Top_k^{\mathrm{indices}}(r_1,\ldots,r_{|\mathcal C(q)|}).
\end{aligned}
\label{eq:rerank}
\end{equation}
Here $g_\phi$ is the vision-language reranker with parameters $\phi$,
$r_j$ is the relevance score of candidate input $v_j$, and
$\Top_k^{\mathrm{indices}}$ returns the indices $I_k(q)$ of the $k$
highest-scoring candidates.
We derive $r_j$ from the yes/no logits and pass those candidates to
the reader with the question.
The reranker input format for the fused retrieval path is detailed
in Appendix~\ref{sec:candidate-details}.

\subsection{Training}
\label{sec:training}

\subsubsection{Contrastive Retrieval Objective}
\label{sec:alignment}
For question $q$, let $e^+$ denote its positive tile and
$\widetilde{\mathcal D}_q$ the training candidate set containing
$e^+$, mined negatives, and in-batch candidates gathered across
devices.
We define the question-side ranking distribution
\begin{equation}
P_q(e)=
\frac{\exp(s_\theta(q,e)/\tau)}
{\sum_{d\in\widetilde{\mathcal D}_q}\exp(s_\theta(q,d)/\tau)},
\label{eq:distribution}
\end{equation}
where $s_\theta(q,e)$ is the encoder similarity between the question
and the fused image--text representation of candidate $e$, and $\tau$
is the softmax temperature.
Following standard contrastive retrieval training, we minimise
\begin{equation}
\mathcal L_{\mathrm{NCE}}=-\log P_q(e^+).
\label{eq:nce}
\end{equation}

\subsubsection{Evidence-Guided Alignment}
\label{sec:evidence-align}
The contrastive term above only uses the question side and does not
exploit evidence supervision.
Let $t_q$ be the evidence supervision text for question $q$ from
Section~\ref{sec:evidence}, and let
$P_t=P_\theta(\cdot\mid t_q)$ be the ranking distribution induced by
$t_q$ over the same candidate set.
We align the question-side ranking with this evidence-side ranking by
\begin{equation}
\mathcal L_{\mathrm{align}}
 =D_{\mathrm{KL}}\bigl(\sg[P_t]\,\|\,P_q\bigr),
\label{eq:align}
\end{equation}
where $D_{\mathrm{KL}}$ is Kullback--Leibler divergence and $\sg$
stops gradients through $P_t$, so that $P_q$ is updated toward the
evidence-induced target.
This makes the evidence text act as a target for the question-side
ranking, which is the side used at inference time.

\subsubsection{Overall Training Objective}
\label{sec:objective}
The total objective combines the contrastive and alignment terms:
\begin{equation}
\mathcal L
 =\mathcal L_{\mathrm{NCE}}+\lambda\,\mathcal L_{\mathrm{align}}.
\label{eq:loss}
\end{equation}
$\lambda$ balances the two terms; we set $\lambda=0.05$ in all
experiments, and both distributions use $\tau\approx0.07$.
Our training signal comes from descriptions of answer evidence in
fixed screenshot tiles.
We adapt the ranking alignment objective of ReAlign \citep{realign}
to these tile-level descriptions rather than selected visual regions.
The adapted encoder builds the fused image--text index used in the
main retrieval pipeline.

\begin{table}[t]
\centering
\small
\begin{tabular}{@{}lcc@{}}
\toprule
Alignment & Weight $\lambda$ & QA \\
\midrule
Contrastive only & -- & 0.7725 \\
Forward KL & 0.20 & 0.7825 \\
Reverse KL & 0.20 & 0.7800 \\
Description (cosine) & 0.10 & 0.7750 \\
Reverse KL & 0.05 & 0.7750 \\
Description (cosine) & 0.20 & 0.7725 \\
\textbf{Forward KL} & \textbf{0.05} & \textbf{0.7900} \\
\bottomrule
\end{tabular}
\caption{Ablation on evidence-alignment objectives. Weight $\lambda$ is
 the coefficient of the alignment term in Eq.~\ref{eq:loss}.
 ``Description (cosine)'' applies a cosine penalty between question and
 description embeddings.}
\label{tab:align-style}
\end{table}

\section{Experiments}
\label{sec:experiments}
\subsection{Evaluation Setting}
\textbf{Benchmarks.}
We evaluate Natural Questions (NQ) \citep{nq}, NQ-Tables,
SimpleQA \citep{simpleqa}, MMSearch \citep{mmsearch}, and
Encyclopedic-VQA (EVQA) \citep{evqa} on Wikipedia QA.
The main comparison reports answer accuracy and gold-article evidence
recall; MMSearch has no gold articles for recall evaluation.

\textbf{Baselines.}
We compare text extraction with mwparserfromhell and Trafilatura
\citep{trafilatura}, screenshot retrieval with PixelRAG (base) and
PixelRAG \citep{pixelrag}, and Pix2Evidence with and without reranking.
Table~\ref{tab:main} presents the main comparison.

\textbf{Configuration.}
The main experiment uses Qwen3-VL-4B-Instruct as the reader and GPT-4.1 \citep{gpt41}
to judge answer accuracy.
Scores use a 0--100 scale unless stated otherwise.
NQ and NQ-Tables are evaluated with either LLM-judged answer
accuracy or exact match (EM), depending on the experiment.
We identify the measure used in each table and figure; further
protocol details appear in Appendix~\ref{sec:evaluation-details}.

\begin{table}[t]
\centering
\small

\setlength{\tabcolsep}{4pt}
\renewcommand{\arraystretch}{1.08}
\begin{tabular*}{\columnwidth}{@{\extracolsep{\fill}}llr@{}}
\toprule
Embedding input & Reader input & SimpleQA \\
\midrule
Image + Text & Image + Text & \textbf{73.4} \\
\midrule
Image + Text & Image & 67.9 \\
Image + Text & Text & 71.9 \\
Text & Image & 60.6 \\
Text & Image + Text & 66.2 \\
Image & Image & 69.0 \\
\bottomrule
\end{tabular*}
\caption{Retrieval--reader modality ablation on SimpleQA. We vary the
 input modality at both stages. Bold indicates the highest score.}
\label{tab:modality-ablation}
\end{table}

\subsection{Main Results}
Table~\ref{tab:main} presents the main comparison.
Without reranking, Pix2Evidence already achieves better results than
both text-extraction and screenshot-retrieval baselines on most
benchmarks, especially on MMSearch and EVQA, reaching 26.0 and 40.6,
respectively.
This shows that fusing visual and textual information in a single
retrieval view provides stronger evidence than using either modality
alone.

Adding the vision-language reranker further improves answer accuracy
on NQ, NQ-Tables, and SimpleQA, reaching 33.2, 27.1, and 77.6,
respectively.
Compared with PixelRAG, these gains are 2.1, 1.3, and 4.8 percentage
points.
The reranked configuration also raises SimpleQA evidence recall from
83.1 to 86.9 and NQ-Tables recall from 50.7 to 55.8.
The non-reranked configuration remains strongest on MMSearch at 26.0
and EVQA at 40.6, which allows the reranking stage to be enabled or
disabled according to the target scenario.

\begin{figure*}[t]
\centering
\includegraphics[width=\textwidth]{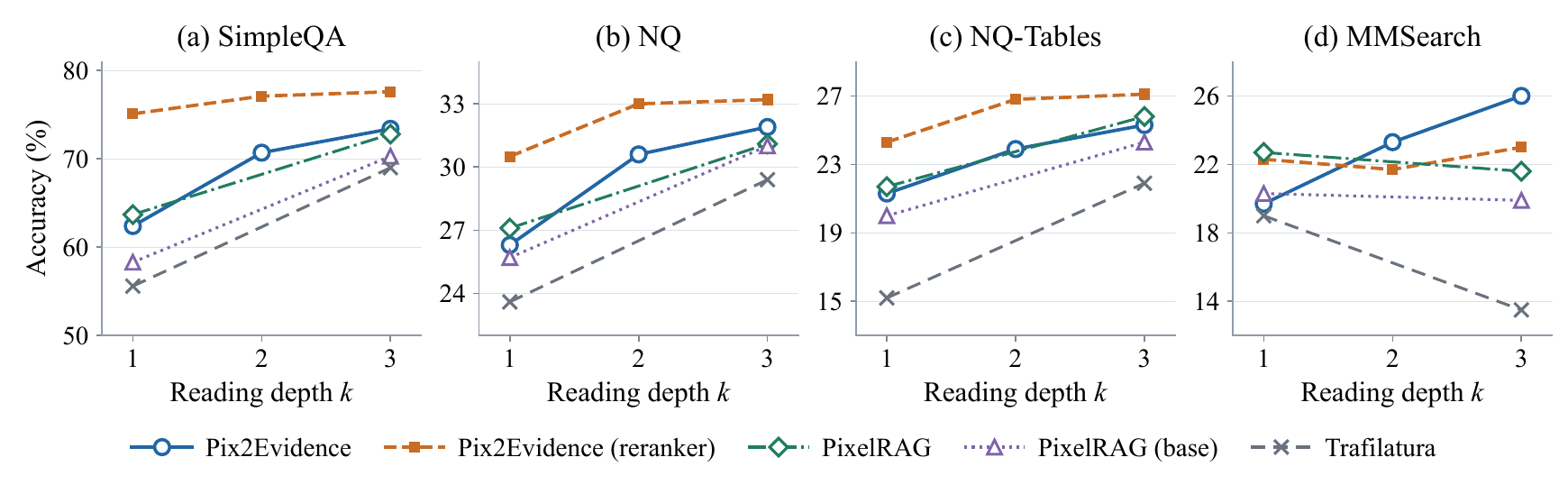}
\caption{Effect of reading depth and reranking across four benchmarks.
 Pix2Evidence and Pix2Evidence (reranker) are evaluated at
 $k\in\{1,2,3\}$. NQ and NQ-Tables use EM in the follow-up runs;
 $k$ denotes the number of items supplied to the reader.}
\label{fig:reading-depth}
\end{figure*}

\begin{table*}[t]
\centering
\small

\setlength{\tabcolsep}{5pt}
\renewcommand{\arraystretch}{1.08}
\begin{tabular*}{\textwidth}{@{\extracolsep{\fill}}lrrrrr@{}}
\toprule
Fusion & SimpleQA & EVQA & NQ (EM) & NQ-Tables (EM) & MMSearch \\
\midrule
Max & 64.6 & 30.4 & 29.9 & 21.7 & 19.0 \\
RRF & 63.7 & 30.4 & \textbf{31.6} & \textbf{24.4} & \textbf{22.0} \\
z-score & \textbf{67.3} & \textbf{33.5} & 29.8 & \textbf{24.4} & 20.3 \\
Min--max & 66.0 & 30.8 & 30.7 & 24.2 & 20.3 \\
\bottomrule
\end{tabular*}
\caption{Fusion-rule ablation for a dual-branch extension without
 reranking ($k=3$). The two branches are an image--text retriever and a
 text-only retriever, with candidates deduplicated by tile. NQ and NQ-Tables use EM. Bold marks column-wise best scores, including ties.}
\label{tab:fusion-ablation}
\end{table*}

\subsection{Evidence-Alignment Variants}
\label{sec:analysis}
Table~\ref{tab:align-style} examines how we train with
teacher-generated evidence descriptions. Following PixelRAG \citep{pixelrag}, we use a development subset for evaluation.
Contrastive only uses $\mathcal L_{\mathrm{NCE}}$.
Forward KL aligns the question-side ranking $P_q$ to the
evidence-side ranking $P_t$ via
$D_{\mathrm{KL}}(\sg[P_t]\,\|\,P_q)$,
which is the direction used in our method, while reverse KL uses the
opposite direction.
Description alignment instead applies a cosine penalty between the
question and description embeddings.
The weight $\lambda$ is the coefficient of the alignment term in
$\mathcal L=\mathcal L_{\mathrm{NCE}}+\lambda\mathcal L_{\mathrm{align}}$
(Eq.~\ref{eq:loss}).

Among the recorded variants, forward KL with $\lambda=0.05$ achieves
the highest QA score, 0.7900, versus 0.7725 for contrastive only.
It also exceeds forward KL at $\lambda=0.20$ (0.7825) and reverse KL
at $\lambda=0.05$ (0.7750).
We therefore adopt forward KL with $\lambda=0.05$ in the final system.

\subsection{Retrieval--Reader Modality Ablation}
\label{sec:modality-ablation}
Table~\ref{tab:modality-ablation} varies the document embedding and
reader inputs on SimpleQA; Text denotes corpus transcriptions.
The best configuration uses Image+Text at both stages, reaching 73.4.

We then fix one stage and vary the other.
With the embedding fixed to Image+Text, the multimodal reader scores
73.4, compared with 71.9 for text-only reading and 67.9 for image-only
reading.
With the multimodal reader fixed, joint embedding scores 73.4,
compared with 66.2 for text-only embedding.
These comparisons show that combining visual and textual information
benefits both retrieval and reading.

\subsection{Dual-Branch Fusion Ablation}
\label{sec:fusion-ablation}
Our main pipeline uses a single fused image--text retrieval view.
As an extension, we analyze how a dual-branch design would behave if a
text-only branch were added, and compare four fusion rules without
reranking at $k=3$ (Table~\ref{tab:fusion-ablation}).
All variants use the same image--text and text branches, with
candidates deduplicated by tile.
Max keeps the larger raw score for each tile; RRF sums reciprocal ranks
with offset 60 \citep{rrf}; z-score and min--max normalise each branch's
scores within a query before summation.

Fusion changes how the two retrieval views prioritise shared
candidates.
Relative to Max, z-score raises SimpleQA from 64.6 to 67.3 and EVQA
from 30.4 to 33.5.
RRF achieves the highest NQ EM at 31.6 and the highest MMSearch score
at 22.0, and ties z-score on NQ-Tables EM at 24.4.
No single rule dominates across all benchmarks, but every fusion rule
outperforms Max on at least one benchmark, showing that combining
normalised scores or branch rankings is more effective than taking the
maximum raw score.

\subsection{Effect of Reading Depth}
\label{sec:reading-depth}
Figure~\ref{fig:reading-depth} varies $k$ from one to three and compares
both Pix2Evidence configurations with PixelRAG, PixelRAG (base), and
Trafilatura at different reading depth.

Across all evaluated depths, both Pix2Evidence configurations
consistently outperform the baselines, and the margin is often large.
For example, on MMSearch, the non-reranked Pix2Evidence reaches 26.0
at $k=3$, compared with 21.6 for PixelRAG, a gain of 4.4 points.
The reranked configuration also reaches 23.0, still above the baseline.
This advantage remains stable across the evaluated reading depths.

\begin{table*}[t]
\centering
\small
\setlength{\tabcolsep}{5pt}
\renewcommand{\arraystretch}{1.10}
\begin{tabular*}{\textwidth}{@{\extracolsep{\fill}}llrrrrr@{}}
\toprule
Reader & Params & SimpleQA & EVQA & NQ (EM) & NQ-Tables (EM) & MMSearch \\
\midrule
\multicolumn{7}{@{}l}{\textit{Qwen3-VL family}} \\
\quad Qwen3-VL-2B & 2.3B & 0.700 & 0.324 & 0.244 & 0.251 & 0.213 \\
\quad Qwen3-VL-4B & 4B & 0.776 & 0.403 & 0.332 & 0.271 & 0.230 \\
\quad Qwen3-VL-8B & 8B & 0.794 & 0.433 & 0.314 & 0.289 & 0.263 \\
\midrule
\multicolumn{7}{@{}l}{\textit{Qwen3.5 family}} \\
\quad Qwen3.5-2B & 2B & 0.752 & 0.421 & 0.235 & 0.229 & 0.213 \\
\quad Qwen3.5-4B & 4B & 0.798 & 0.435 & \textbf{0.335} & 0.271 & 0.273 \\
\quad Qwen3.5-9B & 9B & \textbf{0.813} & \textbf{0.467} & 0.272 & \textbf{0.305} & \textbf{0.300} \\
\bottomrule
\end{tabular*}
\caption{Reader-model ablation with single-branch image--text
 retrieval and vision-language reranking ($N=5$, $k=3$). Values are on a 0--1 scale; NQ and NQ-Tables use EM.
 Bold marks each column's observed maximum.}
\label{tab:reader-ablation}
\end{table*}

\begin{table*}[t]
\centering
\small
\setlength{\tabcolsep}{5pt}
\renewcommand{\arraystretch}{1.08}
\begin{tabular*}{\textwidth}{@{\extracolsep{\fill}}llrrrr@{}}
\toprule
Setting & Pixel fraction & SimpleQA & NQ (EM) & NQ-Tables (EM) & MMSearch \\
\midrule
Baseline & $1$ & 77.6 & 33.2 & 27.1 & 23.0 \\
\midrule
Downsample $c=2$ & $1/2$ & 76.4 & 33.3 & 26.5 & 22.0 \\
Downsample $c=3$ & $1/3$ & 74.4 & 33.4 & 27.7 & 24.0 \\
Downsample $c=4$ & $1/4$ & 74.0 & 33.3 & 26.8 & 20.3 \\
\bottomrule
\end{tabular*}
\caption{Image downsampling on the single-branch Pix2Evidence (reranker)
 configuration ($N=5$, $k=3$), without compression-specific fine-tuning.
 Text is unchanged.}
\label{tab:input-budget}
\end{table*}

\subsection{Reader Model Ablation}
\label{sec:reader-ablation}
Table~\ref{tab:reader-ablation} compares readers from the Qwen3-VL \citep{qwen3vl} and
Qwen3.5 \citep{qwen3.5} families under the same single-branch image--text retrieval
and vision-language reranking setting.
The Qwen3.5 family performs better overall than Qwen3-VL at comparable
or smaller parameter scales.
Qwen3.5-9B gives the best observed results on four benchmarks:
SimpleQA at 0.813, EVQA at 0.467, NQ-Tables at 0.305, and MMSearch at
0.300.
Qwen3.5-4B achieves the highest NQ at 0.335 and also reaches 0.798 on
SimpleQA.
These results indicate that stronger readers generally bring greater
benefits to the system.
The system remains effective across all evaluated reader sizes, and
its performance generally improves as the reader becomes stronger.

\subsection{Image Downsampling}
\label{sec:input-budget}
Table~\ref{tab:input-budget} evaluates image downsampling. We apply Lanczos downsampling, which reduces each image side by
$\sqrt{c}$ and its nominal area by a factor of $c$. Text is kept unchanged.
Additional text-length, candidate-pool, and reading-depth results
appear in Appendix~\ref{sec:budget-extensions}.

Moderate downsampling largely preserves answer quality.
At $c=2$, SimpleQA only drops from 77.6 to 76.4, while NQ slightly
improves from 33.2 to 33.3 and NQ-Tables changes from 27.1 to 26.5.
Stronger compression at $c=4$ reduces SimpleQA to 74.0 and MMSearch to
20.3, while NQ remains stable at 33.3.
Overall, these results show that moderate image downsampling generally
retains most of the measured answer quality with fewer input pixels,
and that SimpleQA and MMSearch are more sensitive to stronger
compression than NQ and NQ-Tables in these runs.

\section{Conclusion}
We presented Pix2Evidence, a multimodal RAG system for question
answering over Wikipedia pages.
It treats each page as a collection of screenshot tiles and encodes
each tile together with its available transcription into a fused
image--text representation, so that layout and readable text are
considered jointly rather than as separate retrieval views.
At inference time, the fused retriever returns candidate tiles, a
vision-language reranker selects the most relevant ones, and a reader
generates the answer from the question and the selected evidence.
For training, a teacher model produces descriptions of
answer-supporting evidence, which guide the retriever through a
contrastive objective and an evidence-alignment term.
Across five Wikipedia QA benchmarks, the non-reranked system scores above
PixelRAG in answer accuracy and leads the main comparison on four.
Its 26.0\% MMSearch accuracy exceeds PixelRAG by 4.4 percentage
points.
The results underline a practical distinction: retaining rich document
content is useful, but a RAG system must also deliver the right
evidence to the reader.
Extensive experiments demonstrate the effectiveness of Pix2Evidence.

\section*{Limitations}
Our evaluation focuses on English Wikipedia QA and a limited set of
model configurations.
Although we test readers from two model families, the sweep does not
cover other architectures, larger scales, or non-English domains.
A broader reader-model evaluation, including multilingual and
domain-specific settings, remains future work. The fused image--text index and the vision-language reranker introduce
additional storage and computation compared with a text-only pipeline.
Our input-budget study varies pixel area and text length, but it does
not measure peak memory and end-to-end latency.
These factors are important for practical deployment, especially when
the reader context is large. A complete cost evaluation and more efficient indexing and reranking
designs therefore remain future work.
For example, one could explore compressed index representations,
lightweight rerankers, or adaptive reranking that is applied only when
the initial retrieval is uncertain.

\section*{Ethical Considerations}
We adhere to the ethical guidelines of the research community.
All experiments use publicly available datasets and open-source models,
ensuring transparency and reproducibility.
We have not collected new data involving human subjects or personally
identifiable information.

The datastore contains rendered web pages and may include copyrighted
material or personal information.
Retrieved evidence improves traceability, but it does not establish
that a source is correct or appropriate to disclose.
Reference answers condition training descriptions only and are not
provided during test-time retrieval or generation.
Our method focuses on retrieval and evidence selection and does not
modify the underlying safety alignment of the reader models.

\bibliography{custom}

\newpage

\appendix
\input{appendix_reorganized}

\end{document}

%% file: appendix_reorganized.tex

\definecolor{pbFrame}{HTML}{7A4A2B}
\definecolor{pbHead}{HTML}{B08D57}
\definecolor{pbBg}{HTML}{FCF8F1}
\definecolor{pbSub}{HTML}{4A4A4A}
\definecolor{pbLine}{HTML}{8C8C8C}
\newlength{\pbWidth}
\newsavebox{\pbBox}
\newcommand{\pbBar}[2]{%
  \setlength{\fboxsep}{5pt}%
  \colorbox{#1}{\parbox{\dimexpr\pbWidth-2\fboxsep\relax}{%
    \color{white}#2}}\par}
\newcommand{\promptfield}[2]{%
  \par\vspace{4pt}%
  \pbBar{pbSub}{\bfseries\small #1}%
  \vspace{1pt}%
  \setlength{\fboxsep}{4pt}%
  \fcolorbox{pbLine}{white}{%
    \parbox{\dimexpr\pbWidth-2\fboxsep-2\fboxrule\relax}{%
      \ttfamily\small #2}}%
  \par\vspace{5pt}%
  \setlength{\fboxsep}{4pt}}
\newenvironment{promptblock}[1]{%
  \par\medskip\noindent
  \setlength{\fboxrule}{0.8pt}%
  \setlength{\fboxsep}{4pt}%
  \begin{lrbox}{\pbBox}%
  \begin{minipage}{\dimexpr\linewidth-2\fboxsep-2\fboxrule\relax}%
    \setlength{\pbWidth}{\linewidth}%
    \pbBar{pbHead}{\bfseries\small #1}%
    \vspace{3pt}%
    \small\raggedright
}{%
  \end{minipage}%
  \end{lrbox}%
  \noindent\fcolorbox{pbFrame}{pbBg}{\usebox{\pbBox}}%
  \par\medskip}

\section{System and Data Construction}
\label{sec:corpus-details}
This appendix distinguishes the PixelRAG screenshot source
\citep{pixelrag} from our transcription, evidence supervision, and
retrieval implementation.
The corpus counts below describe the subset indexed for Pix2Evidence,
not a new rendering effort.

\subsection{Rendered Wikipedia Corpus}
\textbf{Screenshot source.}
We use the Wikipedia images already rendered and tiled by PixelRAG
\citep{pixelrag}, rather than rerendering the pages ourselves.
Its rendering procedure uses an 875-pixel viewport width and divides
long pages into non-overlapping tiles of up to 1024 pixels in height.
Each tile retains a link to its source article.
The image corpus used in our index contains 7,123,459 Wikipedia pages
and 28,244,361 tiles (Table~\ref{tab:data}); these counts differ from
those reported for PixelRAG's full corpus.
We will release the exact source snapshot and  our indexed subset together with the indexed data.

\begin{table}[t]
\centering
\small
\begin{tabular}{@{}p{0.61\columnwidth}r@{}}
\toprule
Item & Count \\
\midrule
Wikipedia pages & 7{,}123{,}459 \\
Screenshot tiles & 28{,}244{,}361 \\
Tiles with nonempty transcriptions & 23{,}948{,}959 \\
Training questions & 39{,}482 \\
Training tile occurrences & 118{,}446 \\
Questions with descriptions & 39{,}205 \\
Hard negatives per question & 2 \\
\bottomrule
\end{tabular}
\caption{Corpus and selected training data. Transcription covers the
 indexed tiles; the transcription count includes only nonempty text.
 Training occurrences include one positive and two negatives per
 question.}
\label{tab:data}
\end{table}

\textbf{Counting units.}
All indexed tiles are processed for transcription, but a tile without
visible text may have an empty transcription.
The 23,948,959 figure counts distinct tiles with nonempty text,
approximately 85\% of the indexed corpus, not transcription coverage.
Training tile occurrences instead count one positive and two negatives
per question; they are not distinct tile records.

\subsection{Text Sources and Supervision}
\textbf{Corpus transcription.}
Qwen3-VL-4B-Instruct processes every indexed tile for visible text,
with a 2048-token output limit.
The prompt discourages completion of words cut off at tile boundaries.
A \texttt{[NO TEXT]} response is stored as an empty string.
Transcriptions are cached independently of the question and answer;
when no text is visible, the fused retriever uses an image-only input.
Text is truncated to a fixed character cap before encoding.

\textbf{Training evidence descriptions.}
For each training question, positive tile, and reference answer,
Qwen3-VL-32B-Instruct is prompted to generate $K$ descriptions of
visible answer evidence; $K$ is configurable and the reported run
uses $K=3$.
Generation uses temperature 0.7, a 1024-token output limit, and
images resized to a maximum side of 2048 pixels.
Generated descriptions are available for 39,205 of 39,482 questions.
One is sampled uniformly at each training use; for questions without
generated descriptions, the question and answer are concatenated.
Thus every training question has supervision text.
The answer helps the teacher locate evidence during data construction.
The supervision is used only during training, not at inference.
We apply quality, length, and format checks before sampling the
descriptions.

\subsection{Model Inputs and Data-Production Prompts}
\label{sec:input-prompts}
\textbf{Input roles.}
Table~\ref{tab:prompt-roles} lists the input fields used at each stage.
In Section~\ref{sec:retrieval}, $f_\theta(q)$ denotes the embedding of
a query-template input, and $f_\theta(x_i,o_i)$ denotes a
document-template input containing the tile image and available text,
or the image alone when $o_i$ is empty.
The text-only input $f_\theta(o_i)$ is used only in the text-only
ablation.

\begin{table}[t]
\centering
\small
\begin{tabular}{@{}p{0.30\columnwidth}p{0.59\columnwidth}@{}}
\toprule
Stage & Input fields \\
\midrule
Transcription & Tile image; text-visible transcription instruction \\
Evidence teacher & Training question, positive tile, reference answer \\
Query embedding & Question in query role \\
Document embedding & Tile image and available transcription \\
Reranking & Question and retrieved candidate payload \\
Answer generation & Question and selected evidence items \\
\bottomrule
\end{tabular}
\caption{Inputs used at each stage. The document embedding uses the
 fused image--text input; a text-only document input is used only in
 the text-only ablation.}
\label{tab:prompt-roles}
\end{table}

\begin{table*}[t]
\centering
\small
\setlength{\tabcolsep}{5pt}
\begin{tabular*}{\textwidth}{@{\extracolsep{\fill}}llrl@{}}
\toprule
Benchmark & Task family & Subset size & Main-table measure \\
\midrule
Natural Questions & Text and Wikipedia QA & 1{,}000 & EM, article recall \\
NQ-Tables & Structured Wikipedia QA & 1{,}000 & EM, article recall \\
SimpleQA & Factual Wikipedia QA & 1{,}000 & GPT-4.1 judged accuracy, article recall \\
MMSearch & Multimodal QA & 300 & GPT-4.1 judged accuracy; no article recall \\
Encyclopedic VQA & Image-grounded QA & 749 & GPT-4.1 judged accuracy, article recall \\
\bottomrule
\end{tabular*}
\caption{Wikipedia benchmark subsets used in our evaluation, together with
their task families, subset sizes, and main-table evaluation measures.
All five tasks query our Wikipedia tile index, while NQ and NQ-Tables use
exact match and the remaining benchmarks use GPT-4.1 judged accuracy.}
\label{tab:benchmark-protocol}
\end{table*}

\begin{table}[t]
\centering
\small
\begin{tabular}{@{}p{0.46\columnwidth}p{0.42\columnwidth}@{}}
\toprule
Retrieval path / hit & Reranker input \\
\midrule
Fused path: image hit & Image; text if available \\
Dual-view ablation: image hit & Image only \\
Dual-view ablation: text hit & Text \\
\bottomrule
\end{tabular}
\caption{Reranker inputs. The fused path is the main pipeline; the
 dual-view entries belong to the fusion ablation in
 Section~\ref{sec:fusion-ablation}.}
\label{tab:payload}
\end{table}

\textbf{Data-production prompts.}
Qwen3-VL-4B-Instruct produces the corpus transcription $o_i$ of
Section~\ref{sec:ocr} from the tile image, and
Qwen3-VL-32B-Instruct produces the training supervision $t_q$ of
Section~\ref{sec:evidence} from the positive tile.
Figure~\ref{fig:data-prompts} reproduces both prompts verbatim; line
breaks are added for display only.
For the evidence description, $\{k\}$ is the number of requested
descriptions, set to three in the reported run, and $\{query\}$ is the
training question.
The final line appends $\{answer\_hint\}$, a training-only marker of the
form \texttt{\textbackslash nGolden answer (use only to locate the
evidence; do NOT quote it verbatim as the whole description):} followed
by the reference answer.

\begin{figure*}[t]
\centering
\begin{promptblock}{Transcription prompt}
\promptfield{Image input}{\{screenshot tile\}}
You are a precise OCR engine. Transcribe ALL text that is actually
visible in this image, in natural reading order (top to bottom, left to
right).\par
\textbf{Rules:}\par
1. Transcribe only text that is visibly present. This image is a crop
and may cut through words, sentences, or tables --- do NOT complete
cut-off text and do NOT guess anything that is not shown.\par
2. Do NOT describe images, icons, layout, colors, or formatting.\par
3. Preserve the wording and numbers exactly; keep line breaks between
distinct lines/paragraphs.\par
4. If the image contains no readable text, output exactly: [NO TEXT]\par
Output only the transcribed text, with no preamble or commentary.
\end{promptblock}

\begin{promptblock}{Evidence-description prompt}
\promptfield{Positive tile}{\{screenshot tile\}}
You are given a screenshot tile cropped from a rendered web/document
page, and a question. This tile was retrieved as the evidence that
answers the question. Think step by step about which visible content in
the tile supports the answer, then write \{k\} SELF-CONTAINED evidence
descriptions that each explain --- grounded in what is actually visible
in this tile --- WHY this tile is relevant to and answers the question.
Each description must stand on its own as a faithful summary of the
tile's query-relevant evidence.\par
\textbf{Guidelines:}\par
1. Ground every description in content actually visible in the tile; do
not invent facts, and do not complete text that is cut off at the crop
boundary.\par
2. Name the key evidence (entities, numbers, dates, titles) explicitly
and include the immediate context needed to interpret it; do not clip
numbers or symbols.\par
3. For tables/charts, mention the relevant header/row/column or
legend/axis/unit so the evidence is interpretable.\par
4. Make the \{k\} descriptions genuinely different in wording and
emphasis (different salient evidence or phrasing), not trivial
paraphrases --- this diversity is used for sampling during training.\par
5. Each description should read like a short factual passage that, on
its own, would let a retriever match it to the question.\par
\textbf{Output format} --- a single JSON object, no markdown fence, no
extra text:\par
\{"think": "your brief step-by-step reasoning", "descriptions":
["desc 1", "desc 2", ...]\}\par
\texttt{Question: \{query\}\{answer\_hint\}}
\end{promptblock}
\caption{The two data-production prompts, reproduced verbatim:
 (top) tile transcription for the corpus text view and (bottom)
 question-conditioned evidence descriptions for retriever training.}
\label{fig:data-prompts}
\end{figure*}

\subsection{Retriever Training}
\label{sec:implementation}
\textbf{Optimisation.}
The recorded launch uses batch size 32, gradient-cache chunks of two
\citep{gradcache}, learning rate $10^{-5}$, 20 warm-up steps, and a
cosine schedule capped at 350 steps.
A text warm-up stage uses about 15K question and answer pairs for
50 steps. The selected retriever checkpoint is at step 100. The visual-token limit is 4096.
Three H800 80GB GPUs train the retriever; a fourth serves the reader
through vLLM \citep{vllm}.

\textbf{Alignment objective.}
Question and evidence-description distributions share a temperature
of approximately 0.07.
The forward objective in Section~\ref{sec:objective} weights the KL
term by 0.05 and stops gradients through the question-induced target.
For the reverse-KL variant, the description-induced distribution is
detached and the query branch can still receive gradients.

\textbf{Development subset.}
The development subset, adopted from PixelRAG, is a compact
mini-datastore for embedding ablations, containing 400 questions and
7,426 tiles.
For each question, it includes all tiles from the gold article plus the
top-100 tiles retrieved by the base embedding model, mixing relevant and
irrelevant candidates.
We use this subset for our training alignment ablation.

\subsection{Candidate Processing and Reader Inputs}
\label{sec:candidate-details}
\textbf{Candidate lists.}
Our main fused retrieval path returns a single ranked list and passes
the top-$N$ hits to the reranker.
In the ablation study, we use a dual-view recall of text and fused
image--text to obtain the candidate list, and deduplicate documents
that appear in both views.
This variant is used only for the fusion ablation in
Section~\ref{sec:fusion-ablation}.
The fusion ablation deduplicates by tile and compares Max, RRF,
z-score, and min--max without reranking.
We apply per-query score normalisation before summation.

\section{Evaluation Protocol}
\label{sec:evaluation-details}

\subsection{Benchmark Details}
\textbf{Benchmark groups.}
NQ, NQ-Tables, and SimpleQA mainly test factual answers in text or
structured page content.
MMSearch and Encyclopedic VQA add image-grounded questions.
Table~\ref{tab:benchmark-protocol} lists the benchmarks used in our
evaluation and their subset sizes.
We report these subset sizes.

\textbf{Natural Questions (NQ).}
We use 1,000 questions associated with Wikipedia articles \citep{nq}.
Answers are checked against the available reference aliases.
This is a text-centric task, but the supporting content can include
structured page elements.

\textbf{NQ-Tables.}
We evaluate 1,000 table-oriented Wikipedia questions.
Tables and infoboxes make the presentation of structured facts
relevant to retrieval and reading.

\textbf{SimpleQA.}
We use 1,000 short factual questions \citep{simpleqa}.
We use the Wikipedia datastore to retrieve evidence for them.
The answer-bearing fact can appear in a paragraph, list, infobox,
or table; a retrieved article is not necessarily the correct tile.

\textbf{MMSearch.}
We include all 300 end-to-end examples:
171 include a query image and 129 contain text only.
Although their topics extend beyond Wikipedia, the retrieval backend
in this comparison is the Wikipedia tile index.
MMSearch does not supply gold article URLs for the article-recall
column, which is therefore marked unavailable.

\textbf{Encyclopedic VQA (EVQA).}
We use 749 automatic questions from the landmarks portion of
Encyclopedic VQA \citep{evqa}.
The iNaturalist portion is excluded because some query images are
unavailable in its official release.
This is a visual question answering task grounded in fine-grained
entity information.
The shorthand EVQA in the main table denotes this subset, not every
category of the full benchmark.

\textbf{MoNaCo.}
The separate agent evaluation contains 1,315 multi-hop Wikipedia
questions per recorded run, all completed
(Section~\ref{sec:monaco}) \citep{wolfson2025monaco}.
PixelRAG's reference protocol uses a GPT-5 ReAct controller with a
single search tool, up to 20 search calls per question and five hits
per call \citep{pixelrag}.
MoNaCo Judge-F1  must not be read as the Wikipedia QA accuracy in
Table~\ref{tab:main}.

\begin{table*}[t]
\centering
\small
\setlength{\tabcolsep}{5pt}
\renewcommand{\arraystretch}{1.08}
\begin{tabular*}{\textwidth}{@{\extracolsep{\fill}}lrrr@{}}
\toprule
Transcription cap (chars) & SimpleQA & NQ (EM) & NQ-Tables (EM) \\
\midrule
Baseline & 77.6 & 33.2 & 27.1 \\
\midrule
128 & 72.1 & 29.7 & 24.7 \\
256 & 72.6 & 31.8 & 25.4 \\
512 & 72.8 & 32.3 & 24.7  \\
1024 & 73.9 & 33.3 & 25.3  \\
2048 & 76.7 & 32.9 & 25.4  \\
\bottomrule
\end{tabular*}
\caption{Transcription truncation using the fused image--text retrieval
 view with reranking, native-resolution images, $N=5$, and $k=3$.
 NQ and NQ-Tables report exact match; other reported scores follow
 their respective benchmark protocols.}
\label{tab:text-budget}
\end{table*}

\begin{table*}[t]
\centering
\small
\setlength{\tabcolsep}{5pt}
\renewcommand{\arraystretch}{1.08}
\begin{tabular*}{\textwidth}{@{\extracolsep{\fill}}lrrrrrr@{}}
\toprule
Setting & $N$ & $k$ & SimpleQA & NQ (EM) & NQ-Tables (EM) & MMSearch \\
\midrule
Baseline & 5 & 3 & 77.6 & 33.2 & 27.1 & 23.0 \\
\midrule
Larger pool & 10 & 3 & 80.9 & 34.2 & 27.4 & 25.7 \\
Larger pool & 20 & 3 & 82.0 & 35.0 & 28.1 & 25.0 \\
Larger pool & 50 & 3 & 82.2 & 33.7 & 28.1 & 25.0 \\
\midrule
Longer context & 5 & 5 & 77.7 & 33.6 & 26.4 & 25.3 \\
\bottomrule
\end{tabular*}
\caption{Candidate-pool size and reading-depth extensions for the fused
 image--text retrieval view with reranking.
 $N$ is the number of candidates supplied to the reranker; $k$ is the
 number of items supplied to the reader. Images and text retain the
 baseline input settings. NQ and NQ-Tables report EM.}
\label{tab:budget-extensions}
\end{table*}

\subsection{Scoring and Grading}
\textbf{Answer accuracy.}
The main table uses Qwen3-VL-4B-Instruct as the reader and GPT-4.1 to
judge generated answers, reporting the share judged correct on a
0--100 scale.
PixelRAG's reference grading protocol classifies answers as
CORRECT, INCORRECT, or NOT\_ATTEMPTED and scores only CORRECT as one
\citep{pixelrag}.
For NQ and NQ-Tables, that protocol accepts up to ten gold answer
aliases joined as alternatives.
Its grader configuration specifies temperature zero, seed 42,
and a maximum of 1,000 output tokens.

\textbf{Evidence recall.}
Article recall at reading depth $k$ records whether an item in the
selected top $k$ comes from a gold source article, matched by URL,
not whether the item contains the answer.
The main comparison uses a three-item reader context unless specified
otherwise.
The MMSearch recall column is unavailable because it has no
annotated gold articles in this protocol.

\textbf{Other evaluation scales.}
The later fusion and input-budget studies use exact match (EM) for NQ
and NQ-Tables, without the main table's LLM judging.
Alignment variants report a QA value on a 0--1 scale.
The MoNaCo appendix reports Judge-F1  on a 0--1 scale.
The Qwen3.5 reader batch also uses a 0--1 scale; NQ and NQ-Tables there
are EM-only, while the remaining benchmarks retain their own judging
protocols.
Scores from these different records cannot be compared by simply
subtracting table cells.

\subsection{Baselines and Configuration}
\textbf{Baselines.}
The main comparison includes no retrieval, two text extraction
pipelines (mwparserfromhell and Trafilatura \citep{trafilatura}),
PixelRAG (base) and PixelRAG \citep{pixelrag}, and Pix2Evidence with
and without reranking.
The PixelRAG reference text indexes use 1,024-token chunks to balance
the approximately 875 visual tokens in a screenshot tile; we follow
the same chunk size for the local text baselines.

\textbf{Configuration.}
We evaluate the system mainly with two readers, Qwen3-VL-4B-Instruct and
Qwen3.5-4B, whereas PixelRAG's main table reports a single Qwen3.5-4B
reader.
Our main pipeline uses the fused image--text retriever and passes its
top-$N$ hits to reranking.
The text-only branch and the dual-view retrieval that unions it with the image--text branch are evaluated in the ablation studies rather than in the main configuration.

\begin{table*}[t]
\centering
\small
\setlength{\tabcolsep}{3pt}
\renewcommand{\arraystretch}{1.13}
\begin{tabular*}{\textwidth}{@{\extracolsep{\fill}}llccccc@{}}
\toprule
Retrieval & Reranker ($k=3$) & EVQA & SimpleQA & NQ (EM) & NQ-Tables (EM) & MMSearch \\
\midrule
Fused     & no  & 0.435 & 0.759 & 0.335 & 0.247 & 0.253 \\
Fused     & yes & 0.435 & 0.798 & 0.335 & 0.271 & 0.273 \\
Dual-view & no  & 0.326 & 0.688 & 0.305 & 0.225 & 0.250 \\
Dual-view & yes & 0.399 & 0.810 & 0.355 & 0.298 & 0.273 \\
\bottomrule
\end{tabular*}
\caption{Qwen3.5 reader results at $k=3$ for the four retrieval
 settings. \emph{Fused} is the single image--text view and
 \emph{dual-view} adds the text-only view. Values are on a 0--1 scale.
 NQ and NQ-Tables are EM-only; EVQA, SimpleQA, and MMSearch follow
 their own judging protocols.}
\label{tab:qwen35-k3}
\end{table*}

\begin{table*}[t]
\centering
\small
\setlength{\tabcolsep}{3pt}
\renewcommand{\arraystretch}{1.13}
\begin{tabular*}{\textwidth}{@{\extracolsep{\fill}}llccccc@{}}
\toprule
Retrieval & Reranker ($k=1$) & EVQA & SimpleQA & NQ (EM) & NQ-Tables (EM) & MMSearch \\
\midrule
Fused     & no  & 0.375 & 0.632 & 0.277 & 0.204 & 0.257 \\
Fused     & yes & 0.431 & 0.766 & 0.313 & 0.229 & 0.257 \\
Dual-view & no  & 0.263 & 0.568 & 0.257 & 0.170 & 0.187 \\
Dual-view & yes & 0.377 & 0.784 & 0.316 & 0.252 & 0.263 \\
\bottomrule
\end{tabular*}
\caption{Qwen3.5 reader results at $k=1$ for the same four settings.
 Scale and benchmark-specific metrics follow
 Table~\ref{tab:qwen35-k3}.}
\label{tab:qwen35-k1}
\end{table*}

\section{Additional Experiments}
\label{sec:budget-extensions}

\subsection{Input and Candidate Budgets}
\label{sec:text-budget}
\textbf{Text truncation.}
Table~\ref{tab:text-budget} starts from the full transcription and
truncates it to a fixed number of characters, keeping
native-resolution images, $N=5$ reranker candidates, and $k=3$ reader
items with the fused image--text retrieval view and reranking.
The untruncated run is the baseline, and each remaining row is an
ablation that shortens the text supplied to the encoder.
At 2048 characters the scores stay close to the baseline: SimpleQA is
76.7 versus 77.6 ($-0.9$ points) and NQ EM is 32.9 versus 33.2
($-0.3$ points), while NQ-Tables EM falls from 27.1 to 25.4
($-1.7$ points).
Truncating more aggressively lowers SimpleQA to 73.9 at 1024 characters
and to 72.1--72.8 at 128--512 characters.
NQ and NQ-Tables respond differently: NQ EM is essentially unchanged at
1024 characters ($+0.1$ points) but drops to 32.3 and below at 512
characters and shorter, whereas NQ-Tables EM stays within 24.7--25.4
at every truncation length, so its loss is not recovered by keeping
more text.
Answer quality therefore declines as more text is removed, with the
penalty concentrated in SimpleQA; a mild truncation such as 2048
characters retains most of the measured quality and is a reasonable
setting.

\textbf{Candidate-pool size.}
Table~\ref{tab:budget-extensions} fixes the reader depth at $k=3$
and varies the reranker pool $N$, again with the fused image--text
retrieval view and reranking.
Relative to $N=5$, a pool of 20 raises SimpleQA by 4.4 points to
82.0, NQ EM by 1.8 to 35.0, NQ-Tables EM by 1.0 to 28.1,
and MMSearch by 2.0 to 25.0.
At $N=50$, SimpleQA reaches 82.2, the highest score in this sweep,
but increases only 0.2 points over $N=20$.
NQ EM instead falls from 35.0 to 33.7; NQ-Tables EM stays at 28.1.
MMSearch peaks at $N=10$ with 25.7 and is 25.0 at both larger pools.
The gains against $N=5$ are therefore positive on all four reported
benchmarks, but their size and best tested pool differ.
Scoring 20 or 50 candidates requires four or ten times as many
query--candidate scoring calls as scoring five, assuming one forward
evaluation per pair.
The pool size should therefore be chosen to balance answer quality
against these additional scoring calls.

\begin{table*}[t]
\centering
\small
\setlength{\tabcolsep}{5pt}
\renewcommand{\arraystretch}{1.08}
\begin{tabular*}{\textwidth}{@{\extracolsep{\fill}}llcrrr@{}}
\toprule
Retrieval view(s) & Index & Rerank & Judge-F1 & API cost (USD)
    & GPU time (h) \\
\midrule
Text               & --      & no  & 0.4755 & \$57.43 & 18.2 \\
Image              & base    & no  & 0.4673 & --      & 91.6 \\
Dual-view          & base    & no  & 0.4708 & \$57.26 & 25.2 \\
Dual-view          & base    & yes & 0.4792 & \$57.82 & 24.7 \\
Dual-view          & aligned & no  & \textbf{0.4816} & \$57.44 & 16.3 \\
Dual-view          & aligned & yes & 0.4688 & --      & 18.1 \\
\bottomrule
\end{tabular*}
\caption{MoNaCo results, 1,315 questions per run. \emph{Text} uses the
 transcription index and \emph{image} the base screenshot index;
 \emph{dual-view} adds a text path to either screenshot index, and ours
 is the index that encodes each tile with its transcription. Judge-F1 is
 on a 0--1 scale; bold marks the best value.}
\label{tab:monaco-cost}
\end{table*}

\textbf{Reader context.}
At $N=5$, increasing $k$ from three to five changes SimpleQA by
+0.1 points, NQ EM by +0.4, NQ-Tables EM by $-0.7$, and MMSearch by
+2.3.
This complements the reading-depth study of
Section~\ref{sec:reading-depth}: the reader can benefit from a longer
context, but the change is small and not positive on every benchmark,
whereas each additional item enlarges the reader's input.
A larger $k$ therefore trades more reading cost for a modest change in answer quality.

\subsection{More Ablations with the Qwen3.5 Reader}
\label{sec:qwen35-reader}
\textbf{Relation to the reader sweep.}
The Qwen3.5-4B row for the fused image--text view with reranking at
$k=3$ in Table~\ref{tab:qwen35-k3} is also reported in the main
reader-model comparison (Table~\ref{tab:reader-ablation}).

\textbf{Configurations.}
Tables~\ref{tab:qwen35-k3} and~\ref{tab:qwen35-k1} report Qwen3.5
reader runs at $k=3$ and $k=1$ under four retrieval settings.
\emph{Fused} retrieves through the single view that encodes each tile
together with its available transcription; \emph{dual-view} adds a
second view that indexes the transcriptions as text.
Each setting is run with and without reranking, which uses
Qwen3-VL-Reranker-2B \citep{qwen3vlretrieval} to rescore the retrieved
candidates before the top-$k$ items are passed to the reader.
The four rows therefore differ only in the retrieval views and whether
reranking is applied.

\textbf{Interpretation.}
Without reranking, the fused view is stronger than the dual-view on
every benchmark at $k=3$, and the same holds at $k=1$: at $k=3$,
SimpleQA is 0.759 versus 0.688 and NQ EM is 0.335 versus 0.305, while
MMSearch is 0.253 versus 0.250.
Reranking changes this ordering.
It raises the fused setting from 0.759 to 0.798 on SimpleQA and from
0.247 to 0.271 on NQ-Tables EM, and lifts the dual-view further, from
0.688 to 0.810 on SimpleQA and from 0.225 to 0.298 on NQ-Tables EM.
With reranking, the dual-view therefore becomes the stronger choice on
most benchmarks at $k=3$, leading on SimpleQA (0.810), NQ EM (0.355),
and NQ-Tables EM (0.298), whereas the fused view keeps the best EVQA
score (0.435).
Increasing the reading depth from one item to three also improves both
settings on most benchmarks: for the dual-view with reranking, SimpleQA
rises from 0.784 to 0.810, NQ EM from 0.316 to 0.355, and NQ-Tables EM
from 0.252 to 0.298.

\subsection{Agentic Search on MoNaCo}
\label{sec:monaco}
\textbf{Setting and metric.}
We evaluate our retrieval backends as the search tool of a GPT-5 ReAct
agent on the full MoNaCo benchmark, 1,315 multi-hop Wikipedia questions.
Only the search backend varies across conditions; the agent loop, the
prompt, and the GPT-5 reader are identical, and every run completed all
1,315 questions.
The agent issues up to 15 search queries per question with three results
per query, a shorter budget than the 20 queries with five hits each of
PixelRAG's reference protocol \citep{pixelrag}.
MoNaCo answers are frequently multi-value lists, so we report judge-F1,
computed from an LLM judge that scores each predicted answer against the
reference answers.
Table~\ref{tab:monaco-cost} reports judge-F1 on a 0--1 scale together
with agent API cost and retrieval GPU time.

\textbf{Backends.}
Text-only search uses the transcription index, and image-only search the
base screenshot index.
Dual-view search unions a screenshot path with a second path that
searches the transcriptions as text; the screenshot path uses either the
base index or ours, which encodes each tile together with its
transcription so that a single retrieved item carries both the rendered
evidence and its text.

\textbf{Results.}
Judge-F1 ranges from 0.4673 to 0.4816 across the six runs.
The best score comes from dual-view search on our index without
reranking (0.4816).
At the same retrieval design, our index adds 0.0108 over the base index
(0.4708 to 0.4816), the largest effect in the table.
Text-only search is competitive at 0.4755 and exceeds image-only search
on the base index (0.4673), which matches our expectation that the
transcription view contributes evidence a screenshot-only index can miss;
combining the two views on our index then gives the best result.
Reranking behaves differently on the two indexes: it raises the
base-index dual-view setting from 0.4708 to 0.4792, but it does not help
on our index (0.4816 to 0.4688), so its benefit is largest where
first-stage retrieval is weaker.

\textbf{Cost.}
Agent-side cost varies little across the backends with complete usage:
text-only search and the dual-view runs each accumulate 9.3--9.8M prompt
tokens and cost \$57.26--\$57.82 per run.
Adding the text view therefore leaves the agent's conversation length
essentially unchanged, because every retrieved item remains a screenshot
tile rather than an extracted passage.
Each reranker invocation adds about 1.2--1.4 seconds.
On the retrieval side, our index is cheaper than the base index for the
same design: 16.3 versus 25.2 hours for dual-view, and 18.1 versus 24.7
hours with reranking.